\documentclass[sts,preprint]{imsart}
\RequirePackage[OT1]{fontenc}

\usepackage{amsthm}
\usepackage{amsmath}
\usepackage{natbib}
\RequirePackage[colorlinks,citecolor=blue,urlcolor=blue,filecolor=blue]{hyperref}
\usepackage{graphicx}
\usepackage[colorinlistoftodos]{todonotes}
\usepackage{mathtools}
\usepackage{url}

\startlocaldefs
\numberwithin{equation}{section}
\theoremstyle{plain}

\endlocaldefs

\usepackage{epstopdf}
\AppendGraphicsExtensions{.tif}
\definecolor{bluish}{rgb}{0,0,0.75}
\newcommand{\ins}[1]{{#1}}%

\newcommand\Mycomb[2][n]{\prescript{#1\mkern-0.5mu}{}C_{#2}}

\begin{document}

\begin{frontmatter}

\title{J. B. S. Haldane's Contribution to the Bayes Factor Hypothesis Test}
\runtitle{Haldane's Contribution to the Bayes Factor}

\author{\fnms{Alexander} \snm{Etz}\ead[label=e1]{Etz.Alexander@gmail.com}\thanksref{t1}}
\affiliation{University of California, Irvine}
\address{Correspondence concerning this article should be addressed to: Alexander Etz, University of California Irvine, Department of Cognitive Sciences, Irvine, CA 92697 USA; or Eric-Jan Wagenmakers, University of Amsterdam, Department of Psychology
Nieuwe Achtergracht 129B, 1018 VZ Amsterdam, The Netherlands. Email may be sent to 
    \printead*{e1} 
    or
    \printead*{e2}} 
	\and
\author{\fnms{Eric-Jan} \snm{Wagenmakers}\ead[label=e2]{EJ.Wagenmakers@gmail.com}\thanksref{t2}}
\affiliation{University of Amsterdam}
\thankstext{t1}{AE was supported by the National Science Foundation Graduate Research Fellowship Program \#DGE-1321846.}
\thankstext{t2}{EJW was supported by the ERC grant ``Bayes or Bust''.}

\runauthor{Etz and Wagenmakers}

\begin{abstract}
This article brings attention to some historical developments that gave rise to the Bayes factor for testing a point null hypothesis against a composite alternative. In line with current thinking, we find that the conceptual innovation ---to assign prior mass to a general law--- is due to a series of three articles by Dorothy Wrinch and Sir Harold Jeffreys (1919, 1921, 1923). However, our historical investigation also suggests that in 1932 J. B. S. Haldane made an important contribution to the development of the Bayes factor by proposing the use of a mixture prior comprising a point mass and a continuous probability density. Jeffreys was aware of Haldane's work and it may have inspired him to pursue a more concrete statistical implementation for his conceptual ideas. It thus appears that Haldane may have played a much bigger role in the statistical development of the Bayes factor than has hitherto been assumed.
\end{abstract}

\begin{keyword}[class=MSC]
\kwd[Primary ]{62F03} 
\kwd[; secondary ]{62-03}
\end{keyword}

\begin{keyword}
\kwd{History of Statistics}
\kwd{Induction}
\kwd{Evidence}
\kwd{Sir Harold Jeffreys}
\end{keyword}

\end{frontmatter}

\section{Introduction}
Bayes factors grade the evidence that the data provide for one statistical model over another. As such, they represent ``the primary tool used in Bayesian inference for hypothesis testing and model selection" \citep[p.\ 378]{Berger2006}. 
In addition, Bayes factors can be used for model-averaging \citep{HoetingEtAl1999} and variable selection \citep{BayarriEtAl2012}. Bayes factors are employed across widely different disciplines such as astrophysics \citep{LattimerSteiner2014}, forensics \citep{TaroniEtAl2014}, psychology \citep{Dienes2014}, economics \citep{MalikovEtAl2015} and ecology \citep{CuthillCharlestoninpress}. Moreover, Bayes factors are a topic of active statistical interest (e.g., \citealp{FouskakisEtAl2015,HolmesEtAl2015,SparksEtAl2015}; for a more pessimistic view see \citealp{Robertinpress}). These modern applications and developments arguably find their roots in the work of one man: Sir Harold Jeffreys.

Jeffreys is a towering figure in the history of Bayesian statistics. His early writings, together with his co-author Dorothy Wrinch \citep{WrinchJeffreys1919,WrinchJeffreys1921,WrinchJeffreys1923}, championed the use of probability theory as a means of induction and laid the conceptual groundwork for the development of the Bayes factor. The insights from this work culminated in the monograph \textit{Scientific Inference} \citep{Jeffreys1931}, in which Jeffreys gives thorough treatment to how a scientist can use the laws of inverse probability (now known as Bayesian inference) to ``learn from experience'' (for a review see \citealp{howie2002} and an earlier version of this paper available at \url{http://arxiv.org/abs/1511.08180v2}). 
 
Among many other notable accomplishments, such as the development of prior distributions that are invariant under transformation and his work in geophysics and astronomy, where he discovered that the Earth's core is liquid, Jeffreys is widely recognized as the inventor of the Bayesian significance test, with seminal papers in 1935 and 1936 \citep{Jeffreys1935, Jeffreys1936furthersig}. The centerpiece of these papers is a number, which Jeffreys denotes $K$, that indicates the ratio of posterior to prior odds; much later, Jeffreys's statistical tests would come to be known as \textit{Bayes factors} \citep{good1958}.\footnote{See \citet{Good1988} and \citet{Fienberg2006} for a historical review. The term `Bayes factor' comes from Good, who attributes the introduction of the term to Turing, who simply called it the `factor.'} Once again these works culminated in a comprehensive book, \textit{Theory of Probability} \citep{Jeffreys1939}. 

When the hypotheses in question are simple point hypotheses, the Bayes factor reduces to a likelihood ratio, a method of measuring evidential strength which dates back as far as Johann Lambert in 1760 \citep{lambert2001} and Daniel Bernoulli in 1777 (\citealp{Kendall1961}; see \citealp{edwards1974} for a historical review); C. S. Peirce had specifically called it a measure of `weight of evidence' as far back as 1878 (\citealp{Peirce1878Thermometer}; see \citealp{Good1979}). Alan Turing also independently developed likelihood ratio tests using Bayes' theorem, deriving \textit{decibans} to describe the intensity of the evidence, but this approach was again based on the comparison of simple versus simple hypotheses; for example, Turing used decibans when decrypting the Enigma codes to infer the identity of a given letter in German military communications during World War II \citep[]{Turing19412012}.\footnote{Turing started his Maths Tripos at King's College in 1931, graduated BA in 1934, and was a Fellow of King's College from 1935-1936. Anthony (A. W. F.) Edwards speculates that Turing might have attended some of Jeffreys's lectures while at Cambridge, where he would have learned about details of Bayes' theorem  (Edwards, 2015, personal communication). According to the college's official record of lecture lists, Jeffreys's lectures 'Probability' started in 1935 (or possibly Easter Term 1936), and in the year 1936 they were in the Michaelmas (i.e., Fall) Term. Turing would have had the opportunity of attending them in the Easter Term or the Michaelmas Term in 1936 (Edwards, 2015, personal communication). 
Jack (I. J.) Good has also provided speculation about their potential connection, ``Turing and Jeffreys were both Cambridge dons, so perhaps Turing had seen Jeffreys's use of Bayes factors; but, if he had, the effect on him must have been unconscious for he never mentioned this influence and he was an honest man. He had plenty of time to tell me the influence if he was aware of it, for I was his main statistical assistant for a year'' \citep[p.\ 26]{Good1980}.  Later, in an interview with David Banks, Good remarks that ``Turing might have seen [Wrinch and Jeffreys's] work, but probably he thought of [his likelihood ratio tests] independently'' \citep[p.\ 11]{banks1996}. Of course, Turing could have learned about Bayes' theorem from any of the standard probability books at the time, such as \citet{todhunter1858}, but the potential connection is of interest. For more detail on Turing's work on cryptanalysis see \citet{Zabell2012}.} 
As \citet{Good1979} notes, Jeffreys's Bayes factor approach to testing hypotheses ``is especially `Bayesian' [because] either [hypothesis] is composite'' (p.\ 393).

Jeffreys states that across his career his ``chief interest is in significance tests'' \citep[p.\ 452]{Jeffreys1980}. Moreover, in an (unpublished) interview with Dennis Lindley (DVL) for the Royal Statistical Society on August 25, 1983, when asked ``What do you see as your major contribution to probability and statistics?'' Jeffreys (HJ) replies,

\begin{quote}
HJ: The idea of a significance test, I suppose, putting half the probability into a constant being 0, and distributing the other half over a range of possible values.

DVL: And that's a very early idea in your work. 

HJ: Well, I came on to it gradually. It was certainly before the first edition of `Theory of Probability'. 

DVL: Well, of course, it is related to those ideas you were talking about to us a few minutes ago, with Dorothy Wrinch, where you were  putting a probability on\dots

HJ: Yes, it was there, of course, when the data are counts. It went right back to the beginning.

(``Transcription of a Conversation between Sir Harold Jeffreys and Professor D.V. Lindley'', Exhibit A25, St John's College Library, Papers of Sir Harold Jeffreys)

\end{quote}
That Jeffreys considers his greatest contribution to statistics to be the development of Bayesian significance tests, tests that compare a point null against a distributed (i.e, composite) alternative, is remarkable considering the range of his accomplishments. 

In their influential introductory paper on Bayes factors, \citet{KassRaftery1995} state,
\begin{quote}
In a 1935 paper and in his book \textit{Theory of Probability}, Jeffreys developed a methodology for quantifying the evidence in favor of a scientific theory. The centerpiece was a number, now called the \textit{Bayes factor}, which is the posterior odds of the null hypothesis when the prior probability on the null is one-half. (p.\ 773,\ abstract) 
\end{quote}
Many distinguished statisticians and historians of statistics consider the development of the Bayes factor to be one of Jeffreys's greatest achievements and a landmark contribution to the foundation of Bayesian statistics. In a recent discussion of Jeffreys's contribution to Bayesian inference, \citet{RobertEtAl2009} recall the importance and novelty of Jeffreys's significance tests,
\begin{quote}
If the hypothesis to be tested is $H_0:\ \theta = 0$, against the alternative $H_1$ that is \textit{the aggregate of other possible values [of $\theta$]}, Jeffreys initiates one of the major advances of \textit{Theory of Probability} by rewriting the prior distribution as a mixture of a point mass in $\theta = 0$ and of a generic density $\pi$ on the range of $\theta$\dots This is indeed a stepping stone for Bayesian Statistics in that it explicitly recognizes the need to separate the null hypothesis from the alternative hypothesis within the prior, lest the null hypothesis is not properly weighted once it is accepted.  (p.\ 157) [emphasis original]
\end{quote}
Some commentators on \citet{RobertEtAl2009} shared their sentiment. \citet{lindley2009} remarked that Jeffreys's ``triumph was a general method for the construction of significance tests, putting a concentration of prior probability on the null value\dots and evaluating the posterior probability using what we now call Bayes factors'' (p.\ 184). \citet{kass2009} noted that a ``striking high-level feature of Theory of Probability
is its championing of posterior probabilities of
hypotheses (Bayes factors), which made a huge contribution
to epistemology'' (p.\ 180). Moreover, \citet{Senn2009} is similarly impressed with Jeffreys's innovation in assigning a concentration of probability to the null hypothesis, calling it ``a touch of genius, necessary to rescue the Laplacian formulation [of induction]'' (p.\ 186).\footnote{However, see Senn's recent in-depth discussion at \url{http://tinyurl.com/ow4lahd} for a less enthusiastic perspective.}

In discussions of the development of the Bayes factor, as above, most authors focus on the work of Jeffreys, with some mentioning the early related work by Turing and Good. A piece of history that is missing from these discussions and commentaries is the contribution of John Burdon Sanderson (J. B. S.) Haldane, whose application of these ideas potentially spurred Jeffreys into making his conceptual ideas about scientific learning more concrete---in the form of the Bayes factor.\footnote{\citet[p.\ 125]{howie2002} gives a brief account of some ways Haldane might have influenced Jeffreys's thoughts, but does not draw this connection.} In a paper entitled ``The Bayesian Controversy in Statistical Inference'', after discussing some of Jeffreys's early Bayesian developments, \citet{Barnard1967} briefly remarks, 

\begin{quote}
Another man whose views were closely related to Jeffreys was Haldane, who\dots proposed a prior having a ‘lump’ of probability at the null hypothesis with the rest spread out, in connexion [sic] with tests of significance.  (p.\ 238)
\end{quote}
Similarly, the \textit{Biographical Memoirs of the Fellows of the Royal Society} includes an entry for Haldane \citep{pirie1966JBS}, in which M.\ S.\ Bartlett recalls,
\begin{quote}
In statistics, [Haldane] combined an objective approach to populations with an occasional cautious use of inverse probability methods, the latter being apparently envisaged in frequency terms ... [Haldane's] idea of a concentration of a prior distribution at a particular value was later adopted by Harold Jeffreys, F.R.S. as a basis for a theory of significance tests. (p.\ 233)
\end{quote}
However, we have not seen Haldane's connection to the Bayes factor hypothesis test mentioned in the modern statistics literature, and we are not aware of any in-depth accounts of this particular innovation to date.

Haldane is perhaps best known in the statistics literature by his proposal of a prior distribution suited for estimation of the rate of rare events, which has become known as \textit{Haldane's prior} \citep{Haldane1932,Haldane1948}.\footnote{Interestingly, Haldane's prior appears to be an instance of Stigler's law of eponymy, since Jeffreys derived it in his book \textit{Scientific Inference} \cite[p.\ 194]{Jeffreys1931} eight months before Haldane's publication.} References to Haldane's 1932 paper focus mainly on its proposal of the Haldane prior, and they largely miss his formulation of a mixture prior comprising a point mass and a smoothly distributed alternative---a crucial component in the Bayes factor hypothesis tests that Jeffreys would later develop. %
\ins{Among Haldane's various biographies (e.g., \citealp{Clark1968,crow1992,lai1998,sarkar1992centenary}) there is no mention of this development; while they sometimes mention statistics and mathematics among his broad list of interests, they understandably tend to focus on his major advances made in biology and genetics. In fact, this result is not mentioned even in Haldane's own autobiographical account of his career accomplishments \citep{haldane1966}.}

The primary purpose of this paper is to review the work of Haldane and discuss how it may have spurred Jeffreys into developing his highly influential Bayesian significance tests. We begin by reviewing the developments made by Haldane in his 1932 paper, followed by a review of Jeffreys's earliest work on the topic. We go on to draw parallels between their respective works and speculate on the nature of the connection between the two men.

\section{Haldane's Contribution: A Mixture Prior}
J. B. S. Haldane  was a true polymath; \citet{white1965jbs} called him ``probably the most erudite biologist of his generation, and perhaps of the [twentieth] century'' (as cited in \citealp{crow1992}, p.\ 1). 
Perhaps best known for his pioneering work on mathematical population genetics (alongside Ronald Fisher and Sewall Wright, see \citealp{smith1992jbs}), Haldane is also recognized for making substantial contributions to the fields of physiology, biochemistry, mathematics, cosmology, ethics, religion, and (Marxist) philosophy.\footnote{\citet{haldane1966} provides an abridged autobiography, and \citet{Clark1968} is a more thorough biographical reference. For more details on the wide-reaching legacy of Haldane, see \citet{crow1992}, \citet{lai1998}, and \citep{pirie1966JBS}. Haldane was also a prominent public figure during his time and wrote many essays for the popular press (e.g.,  \citealp{haldane1927possible}).} 
\ins{In addition to this already impressive list of topics, in 1932 Haldane published a paper in which he presents his views regarding the foundations of statistical inference \citep{Haldane1932}. At the time this was unusual for Haldane, as most of his academic work before 1932 was primarily concerned with physical sciences.

By 1931, Haldane had been working for twenty years developing a mathematically rigorous account of the chromosomal theory of genetic inheritance; this work began in 1911 (at age 19) when, during a Zoology seminar at New College, Oxford, he announced his discovery of genetic linkage in vertebrates \citep{haldane1966}. This would become the basis of one of his most influential research lines,
to which he intermittently applied Bayesian analyses.\footnote{In one such analysis, \citet{haldane1919probable} used Bayesian updating of a uniform prior (as was customary in that time) to find the probable error of calculated linkage estimates (proportions). It is unclear from where exactly Haldane learned inverse probability, but over the years he occasionally made references to probabilistic concepts put forth by von Mises (e.g., the idea of a ``kollective'' from \citealp{von1931}) or proofs for his formulas given by \citet{todhunter1865}.} 
Throughout his career, Haldane would also go on to publish many papers pertaining to classical (non-Bayesian) mathematical statistics, including 
a presentation of the exact moments of the $\chi^2$ distribution \citep{Haldane1937}, 
a discussion of how to transform various statistics so that they are approximately normally distributed \citep{Haldane1938}, 
an exploration of the properties of inverse (i.e., negative) binomial sampling \citep{Haldane1945}, 
a proposal for a two-sample rank test \citep{Haldane1947}, 
and an investigation of the bias of maximum likelihood estimates \citep{Haldane1951}. Despite his keen interest in mathematical statistics, Haldane's work pertaining to its foundations are confined to a single paper published in the early 1930s. So, while unfortunate, it is perhaps understandable that the advances he made in 1931 are not widely known: This work is something of an anomaly, buried and forgotten in his sizable corpus.} 

\begin{figure}
\includegraphics[scale=2.1]{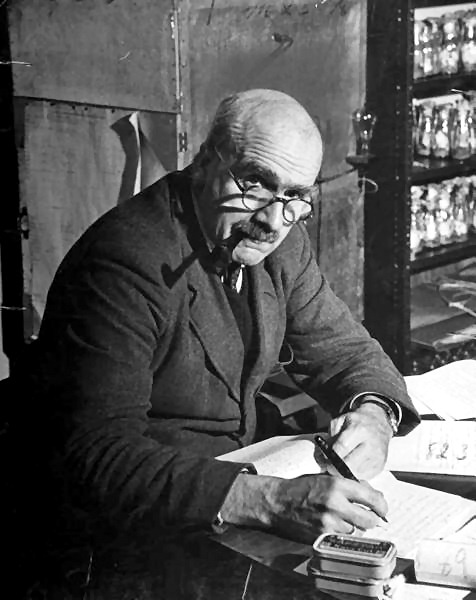}
\caption{John Burdon Sanderson (J. B. S.) Haldane (1892--1964) in 1941. (Photograph by Hans Wild, LIFE Magazine)}
\end{figure}

\ins{Haldane's paper, ``A note on inverse probability'', was received by the \textit{Mathematical Proceedings of the Cambridge Philosophical Society} on November 19, 1931 and read at the Society meeting on December 7, 1931 \citep{Haldane1932}.} He begins 
by stating his goal, ``Bayes' theorem is based on the assumption that all values of the parameter in the neighbourhood of that observed are equally probable \textit{a priori}. It is the purpose of this paper to examine what more reasonable assumption may be made, and how it will affect the estimate based on the observed sample'' (p. 55). 
Haldane frames his paper as giving Fisher's method of maximum likelihood a foundation through inverse probability (much to Fisher's chagrin, see \citealp{Fisher1932}). Haldane gives a short summary of his problem of interest (we will repeatedly quote at length to provide complete context):

\begin{quote}
Let us first consider a population of which a proportion $x$ possess the character $X$, and of which a sample of $n$ members is observed, $n$ being so large that one of the exponential approximations to the binomial probability distribution is valid. Let $a$ be the number of individuals in the sample which possess the character $X$. (p.\ 55)
\end{quote}
The problem, in slightly archaic notation, is the typical binomial sampling setup. 
Haldane goes on to say,
\begin{quote}
It is an important fact that in almost all scientific problems we have a rough idea of the nature of $f(x)$ [the prior distribution] derived from the past study of similar populations. Thus, if we are considering the proportion of females in the human population of any large area, $f(x)$ is quite small unless $x$ lies between .4 and .6. (p.\ 55)
\end{quote}
Haldane appeals to availability of real scientific information to justify using non-uniform prior distributions. It is stated as a fact that we have information indicating that various regions for $x$ are more probable than others. Haldane goes on to derive familiar posteriors for $x$ starting from a uniform prior (p.\ 55-56) before discussing possible implications that deviations from uniform priors have on the posteriors. His description of one such interesting deviation follows:

\begin{quote}
If [the slope of the prior probability density], though small compared with $n^{1/2}$ in the neighborhood of $x=a/n$, has an infinity or becomes very large for some other value of $x$ (other than $0$ or $1$), and if $a/n$ is finite [i.e., $a$ $\neq$ $0$ and $n$ $\neq$ $\infty$], then the [posterior] probability distribution is approximately Gaussian in the neighborhood of $x=a/n$, but has an infinity or secondary maximum at the other point or points. (p.\ 56-57)
\end{quote}
In other words, when a prior distribution has a distinct mass of probability at some point between $0$ and $1$, and a large binomial sample is obtained that contains some $a$'s and some $not$-$a$'s, the posterior distribution can be approximated by a Gaussian except for a separate infinity. 

Haldane avoids the trouble of an infinite density in the prior\footnote{Incidentally, Jeffreys is struggling to solve a similar problem of infinite densities in a prior distribution around this time (see \citealp{Jeffreys1931}, pp.\ 194-195). } 
by marginalizing across two orthogonal models to obtain a single mixture prior distribution that consists of a point hypothesis and a smoothly distributed alternative.
Haldane first clarifies with an example and then solves the problem. We quote at length:

\begin{quote}
An illustration from genetics will make the point clear. The plant \textit{Primula sinensis} possesses twelve pairs of chromosomes of approximately equal size. A pair of genes selected at random will lie on different chromosomes in $11/12$ of all cases, giving a proportion $x=.5$ of ``cross-overs.'' In $1/12$ of all cases they lie on the same chromosome, the values of the cross-over ratio $x$ ranging from $0$ to $.5$ without any very marked preference for any part of this range, except perhaps for a tendency to avoid values very close to $.5$. $f(x)$ is thus approximately $1/6$ for $0$ $\leq$ $x$ $<$.5; it has a discontinuity at $x=.5$, such that the probability is $11/12$; while, for $.5 < x \leq 1$, $f(x)=0$.

\hspace{4mm} Now if a family of $400$ seedlings from the cross between 
[two plants] contains $160$ ``cross-overs'' we have two alternatives. The probability of getting such a family from a plant in which the genes lie in different chromosomes is $11/12$ $\Mycomb[400]{160}$ $2^{-400}$, or $1.185\times 10^{-5}$. The probability of getting it from a plant in which they lie in the same chromosome is
$$\frac{1}{6} \Mycomb[400]{160} \int_{0}^{.5} x^{160}(1-x)^{240}dx.$$
Since this integral is very nearly equal to 
$$\int_{0}^{1} x^{160}(1-x)^{240}dx,\ \textrm{or}\ \frac{160!\ 240!}{401!}, $$
this probability is approximately $\dfrac{1}{6\times 401}$, or $4.56\times 10^{-4}$. Thus the probability that the family is derived from a plant where the genes lie in different chromosomes and $x=.5$ is $.028$. Otherwise the mean value of $x$ is $.4$, with standard error $.0245$. The overall mean value, or mathematical expectation, of $x$ is $.4028$, and the graph of the [posterior probability density] is an approximately normal error curve centred at $x=.4$ with standard deviation $.0245$, together with an infinity at $x=.5$.  (p.\ 57)
\end{quote}
Haldane's passage may be hard to parse since the example is somewhat opaque and the notation is dated. However, the passage is crucial and therefore we unpack Haldane's problem as follows. 
When Haldane speaks of ``pairs of genes'' he means that there are two different genes that are responsible for different traits, such as genes for stem length and leaf color in \textit{Primula sinensis}. Since the DNA code for particular genes are located in specific parts of specific chromosomes, during reproduction they can randomly have their alleles switch from mother chromosome to father chromosome, which is called ``cross-over.'' We are cross-breeding plants and we want to know the cross-over rate for these genes, which depends on whether the pair of genes are on the same of different chromosomes. For example, if the gene for petal color and the gene for stem length are in different chromosomes, then they would cross-over independently in their respective chromosomes during cell division, and the children should show new mixtures of color and length at a certain rate (50\% in Haldane's example). If they are in the same chromosome it is possible for the two genes to be located in the same segment of the chromosome that crosses over, and because their expression varies together they will show less variety in trait combination on average (i.e., $<50\%$). 

Haldane's example uses this fact to go backwards, from the number of ``cross-overs'' present in the child plants to infer the chromosomal distance between the genes. We will use $\theta$, rather than Haldane's $x$, to denote the cross-over rate parameter. If the different genes lie on different chromosomes they cross-over independently during meiosis, and so there is a 50\% probability to see new combinations of these traits for any given offspring. Hence if traits are on different chromosomes then $\theta=.5$. If they lie on the same chromosome they have a cross-over rate of $\theta<.5$, where the percentage varies based on their relative location on the chromosome. If they are relatively close together on the chromosome they are likely to cross-over together and we won't see many offspring with new combinations of traits, so the cross-over rate will be closer to $\theta=0$. If they are relatively far apart on the chromosome they are less likely to cross-over together, so they will have a cross-over rate closer to $\theta=.5$. 

Since there are 12 pairs of chromosomes, there is a natural prior probability assignment for the two competing models: 11/12 pairs of genes selected at random will lie on different chromosomes ($\mathcal M_0$) and 1/12 will lie on the same chromosome ($\mathcal M_1$); when they are on the same chromosome they could be anywhere on the chromosome, so the distance between them can range from nearly nil to nearly an entire chromosome. To capture this information Haldane uses a uniform prior from $0$ to $.5$ for $\theta$. When they are on different chromosomes, $\theta=.5$ precisely. Hence Haldane's mixture prior comprises the prior distributions for $\theta$ from the two models,
$$
\begin{array}{rcl}
\pi_0(\theta)&=&\delta(.5)\\
\pi_1(\theta)&=& \mathcal U(0,.5),
\end{array}
$$ 
where $\delta(\cdot)$ denotes the Dirac delta function, and with prior probabilities (i.e., mixing weights) for the two models of $P(\mathcal M_0)=11/12$ and $P(\mathcal M_1)=1/12$. The marginal prior density for $\theta$ can be written as
$$
\begin{array}{rcl}
\pi(\theta) &=& P(\mathcal M_0)\pi_0(\theta)+P(\mathcal M_1)\pi_1(\theta)\\[3ex]
&=& \dfrac{11}{12}\times\delta(.5) + \dfrac{1}{12}\times \mathcal U(0,.5),
\end{array}
$$
using the law of total probability. Haldane is left with a mixture of a point mass and a smooth probability density.

Haldane breeds his plants and obtains $n = 400$ offspring, $a=160$ of which are cross-overs (an event we denote $D$). The probability of the data, 160 cross-overs out of 400, under $\mathcal{M}_0$ (now known as the marginal likelihood) is 
$$P(D \mid \mathcal{M}_0) = \begin{pmatrix}
400\\
160
\end{pmatrix} (.5)^{400}.$$
The probability of the data under $\mathcal M_1$ is
$$ P(D \mid \mathcal{M}_1) = 2 \, \begin{pmatrix}
400\\
160
\end{pmatrix} \int_{0}^{.5} \theta^{160}(1- \theta)^{240}\,\text{d}\theta .$$
The probabilities of the data can be used in conjunction with the prior model probabilities to update the prior mixing weights to posterior mixing weights (i.e., posterior model probabilities) by applying Bayes' theorem as follows (for $i = 0, 1$):
$$
P(\mathcal{M}_i\mid D)=
\frac{P(\mathcal{M}_i)P(D\mid\mathcal{M}_i)}{P(\mathcal{M}_1)P(D\mid \mathcal{M}_1)+P(\mathcal{M}_0)P(D\mid \mathcal{M}_0)}.
$$
Using the information found thus far, the  posterior model probabilities are $P(\mathcal{M}_1\mid D)= .972$ and $P(\mathcal{M}_0\mid D)= .028$. The two conditional prior distributions for $\theta$ are also updated to conditional posterior distributions using Bayes' theorem. Under $\mathcal M_0$ the prior distribution is a Dirac delta function at $\theta=.5$, which is unchanged by the data. Under $\mathcal M_1$, the prior distribution $\mathcal U(0,.5)$ is updated to a posterior distribution that is approximately $\mathcal N(.4, .0245^2)$.   
The marginal posterior density for $\theta$ can thus be written as a mixture,
$$
\begin{array}{rcl}
\pi(\theta\mid D) &=& P(\mathcal M_0\mid D)\pi_0(\theta\mid D)+P(\mathcal M_1\mid D)\pi_1(\theta\mid D)\\[3ex]
&=& .028\times \delta(.5) + .972\times \mathcal N(.4,.0245^2).
\end{array}
$$
Moreover, Haldane then uses the law of total probability to arrive at a model-averaged prediction for $\theta$, as follows: $E(\theta) = .5 \times .028 + \frac{160}{400} \times .972 = .4028$. This appears to be the first concrete application of \textit{Bayesian model averaging} \citep{HoetingEtAl1999}.\footnote{\citet[p.\ 166]{RobertEtAl2009} point out that Jeffreys's \textit{Theory of Probability} \citep{Jeffreys1939} ``includes the seeds'' of model averaging. In fact, the seeds appear to go back to \citet[p.\ 387]{WrinchJeffreys1921}, where they briefly note that if future  observation $q_2$ is implied by law $p$ (i.e., $P(q_2\mid q_1,p)=1$), ``the probability of a further inference from the law is not appreciably higher than that of the law itself'' since the second term in the sum $P(q_2\mid q_1) = P(p\mid q_1)P(q_2\mid q_1,p)+P(\sim p\mid q_1)P(q_2\mid q_1,\sim p)$ is usually ``the product of two small factors'' (p.\ 387).}

Haldane uses a mixture prior distribution to solve another challenging problem.\footnote{The way \citet{Haldane1932} sets up the problem shows a great concurrence of thought with Jeffreys, who was tackling a similar problem at the time in his book \textit{Scientific Inference} \citep[p.\ 194-195]{Jeffreys1931}.} 
 We quote Haldane:

\begin{quote}
We now come to the case where $a=0$\dots Unless we have reasons to the contrary, we can no longer assume that $x=0$ is an infinitely improbable solution\dots In the case of many classes both of logical and physical objects we can point to a finite, though undeterminable, probability that $x=0$. Thus there are good, but inconclusive, reasons for the beliefs that there is no even number greater than 2 that cannot be expressed as the sum of 2 primes, and that no hydrogen atoms have an atomic weight between 1.9 and 2.1. In addition, we know of very large samples of each containing no members possessing these properties. Let us suppose then, that $k$ is the \textit{a priori} probability that $x=0$, and that the \textit{a priori} probability that it has a positive [nonzero] value is expressed by $f(x)$, where $\underset{\epsilon \rightarrow 0}{Lt}\int_{\epsilon}^1f(x)dx=1-k$. (p.\ 58-59)
\end{quote}
There is an explicit (albeit data-driven, by the sound of it) interest in a special value of the parameter, but under a continuous prior distribution the probability of any point is zero. To solve this problem, Haldane again uses a mixture prior; in modern notation, $k$ denotes the prior probability $P(\mathcal M_0)$ of the point-mass component of the mixture, a Dirac delta function at $\theta=0$, with a second component that is a continuous function of $\theta$ with prior probability $P(\mathcal M_1)=1-k$.

In a straightforward application of Bayes' theorem, Haldane finds that 
``the probability, after observing the sample, that $x=0$ is
$$\dfrac{k}{k+\int_0^1(1-x)^nf(x)dx}. 
\hspace{3mm}$$
``If $f(x)$ is constant this is $(n+1)k/(nk+1)$\dots This is so even if $f(x)$ has a logarithmic infinity at $x=0$... Hence as $n$ tends to infinity the probability that $x = 0$ tends to unity, however small be the value of $k$.'' \citep[p.\ 59]{Haldane1932}. Haldane again goes on to perform Bayesian model-averaging to find ``the probability that the next individual observed will have the character $X$'' (p.\ 59).

In sum, in 1931 Haldane presents his views on the foundations of statistical inference, and  proposed to use a two-component mixture prior comprising a point mass and smoothly distributed alternative. He went on to apply this mixture prior to concrete problems involving genetic linkage, and in doing so also performed the first instances of Bayesian model-averaging. To assess the importance and originality of Haldane's contribution it is essential to discuss the related earlier work of Wrinch and Jeffreys, and the later work by Jeffreys alone.

\section{Wrinch and Jeffreys's Development of the Bayes Factor before Haldane (1932)}
Jeffreys was interested in the philosophy of science and induction from the beginning of his career. He learned of inverse probability at a young age (circa 1905, when he would have been 14 or 15 years old) from reading his father's copy of \citet{todhunter1858} (Exhibit H204, St John's College Library, Papers of Sir Harold Jeffreys). To Jeffreys, Toddhunter explained ``inverse probability
absolutely clearly and I [Jeffreys] never saw there was any difficulty about it'' (Transcribed by AE from the audio-cassette in Exhibit H204, St John's College Library, Papers of Sir Harold Jeffreys). His views were refined around the year 1915 while studying Karl Pearson's influential \textit{Grammar of Science} \citep{pearson1892}, which led Jeffreys to see inverse probability as the method that ``seemed\dots the sensible way of expressing common sense'' (``Transcription of a Conversation between Sir Harold Jeffreys and Professor D.V. Lindley'', Exhibit A25, St John's College Library, Papers of Sir Harold Jeffreys). This became a theme that permeated his early work, done in conjunction with Dorothy Wrinch, which sought to put probability theory on a firm footing for use in scientific induction.\footnote{More background for the material in this section can be found in \citet{Aldrich2005} and \citet{howie2002}. See \citet[Chapter 4]{howie2002} and \citet[Chapter 8]{senechal2012died} for details on the personal relationship between Wrinch and Jeffreys. Many of the ideas presented by Wrinch and Jeffreys are similar in spirit to those of W. E. Johnson; as a student, Wrinch attended Johnson's early lectures on advanced logic at Cambridge (see \citealp[p.\ 86]{howie2002}). Jeffreys would later emphasize Wrinch's important contributions their work on scientific induction, ``I should like to put on record my appreciation of the substantial contribution she [Wrinch] made to this work [\citep{WrinchJeffreys1919,WrinchJeffreys1921,WrinchJeffreys1923,WrinchJeffreys1923mensuration}], which is the basis of all my later work on scientific inference'' \citep[p.\ 564]{HodgkinJeffreys1976}. See \citet{senechal2012died} for more on Wrinch's far-reaching scientific influence.}

\begin{figure}
\includegraphics[scale=2]{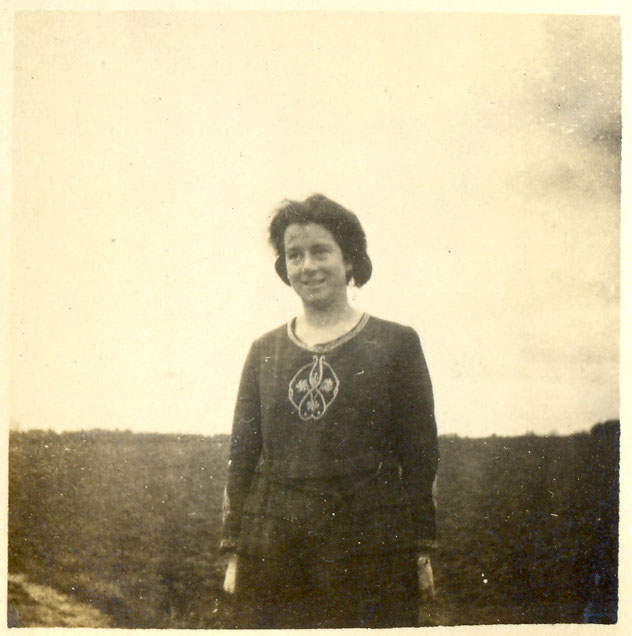}
\caption{Dorothy Wrinch (1894--1976) circa 1920--1923. (Photograph by Sir Harold Jeffreys, included by permission of the Master and Fellows of St John's College, Cambridge)}
\end{figure}

Their work was motivated by that of \citet{Broad1918}, who showed\footnote{For an in-depth discussion about the law of succession, including Broad's insights, see \citet{Zabell1989}. Zabell points out that Broad's result had been derived earlier by others, including \citet{Prevost1799}, \citet{Ostrogradski1848}, and \citet{Terrot1853}. Interested readers should see \citet[p.\ 286, as well as the mathematical appendix beginning on p.\ 309]{Zabell1989}.}
that when one applies Laplace's principle of insufficient reason --- assigning equal probability to all possible states of nature --- to finite populations, one is lead to an inductive pathology: A general law could virtually never achieve a high probability. This was a result that flew in the face of the common scientific view of the time that ``an
inference drawn from a simple scientific law may have a very high probability, not far from
unity'' \citep[p.\ 380]{WrinchJeffreys1921}. \citet{Jeffreys1980} later recalled Broad's result,

\begin{quote}
Broad used Laplace's theory of sampling, which supposes that if we have a population of $n$ members, $r$ of which may have a property $\varphi$, and we do not know $r$, the prior probability of any particular value of $r$ (0 to $n$) is 1/($n$+1). Broad showed that on this assessment, if we take a sample of number $m$ and find all of them with $\varphi$, the posterior probability that all $n$ are $\varphi$'s is ($m$+1)/($n$+1). A general rule would never acquire a high probability until nearly the whole of the class had been sampled. We could never be reasonably sure that apple trees would always bear apples (if anything). The result is preposterous, and started the work of Wrinch and myself in 1919--1923. (p.\ 452)
\end{quote}

The consequence of Wrinch and Jeffreys's series of papers was the derivation of two key results. First, they found a solution to Broad's quandary in the assignment of a finite initial probability, independent of the population's size, to the general law itself, which allows a general law to achieve a high probability without needing to go so far as to sample nearly the entire population.  
Second,  they derived the odds form of Bayes' theorem \citep[p.\ 387]{WrinchJeffreys1921}: ``If $p$ denote the most probable law at any stage, and $q$ an additional experimental fact,
[and $h$ the knowledge we had before the experiment,]'' (p.\ 386) a new form of Bayes' theorem can be written as
$$
\frac{P(p \mid q.h)}{P(\sim p \mid q.h)} = 
\frac{P(q \mid p.h)}{P(q \mid \sim p.h)} \cdot
\frac{P(p \mid h)}{P(\sim p \mid h)}. 
$$
At the time, the conception of Bayes' rule in terms of odds was  novel. \citet{Good1988} remarked, ``[the above equation] has been mentioned several times in the literature without citing Wrinch and Jeffreys. Because it is so important I think proper credit should be given'' (p.\ 390).
We agree with Good; the importance of this innovation should not be understated, because it lays the foundation for the future of Bayesian hypothesis testing. 
A simple rearrangement of terms highlights that the Bayes factor is the ratio of posterior odds to prior odds,
$$ 
\underbrace{\frac{P(q \mid p,h)}{P(q \mid \sim p,h)}}_\text{Bayes factor} =
\underbrace{\frac{P(p \mid q,h)}{P(\sim p \mid q,h)}}_\text{Posterior odds} \bigg/
\underbrace{\frac{P(p \mid h)}{P(\sim p \mid h)}}_\text{Prior odds}.
$$
Wrinch and Jeffreys went on to show how the Bayes factor---which is the amount by which the data, $q$, shift the balance of probabilities for $p$ versus $\sim p$---can form the basis of a philosophy of scientific learning \citep{LyEtAlinpress}. 
\section{Jeffreys's Development of the Bayes Factor after Haldane (1932)}
Remarkably, the Bayes factor remained a conceptual development until Jeffreys published two seminal papers in 1935 and 1936. In 1935, Jeffreys published his first significance tests in the \textit{Mathematical Proceedings of the Cambridge Philosophical Society,} with the title, ``Some Tests of Significance, Treated by the Theory of Probability'' \citep{Jeffreys1935}. This was shortly after his published dispute with Fisher (for an account of the controversy see \citealp{Aldrich2005,howie2002,Lane1980}). 

In his opening statement to the 1935 article, Jeffreys states his main goal: 
\begin{quote}
It often happens that when two sets of data obtained by observation give slightly different estimates of the true value we wish to know whether the difference is significant. The usual procedure is to say that it is significant if it exceeds a certain rather arbitrary multiple of the standard error; but this is not very satisfactory, and it seems worth while to see whether any precise criterion can be obtained by a thorough application of the theory of probability. (p.\ 203)
\end{quote}
\begin{figure}
\includegraphics[scale=1]{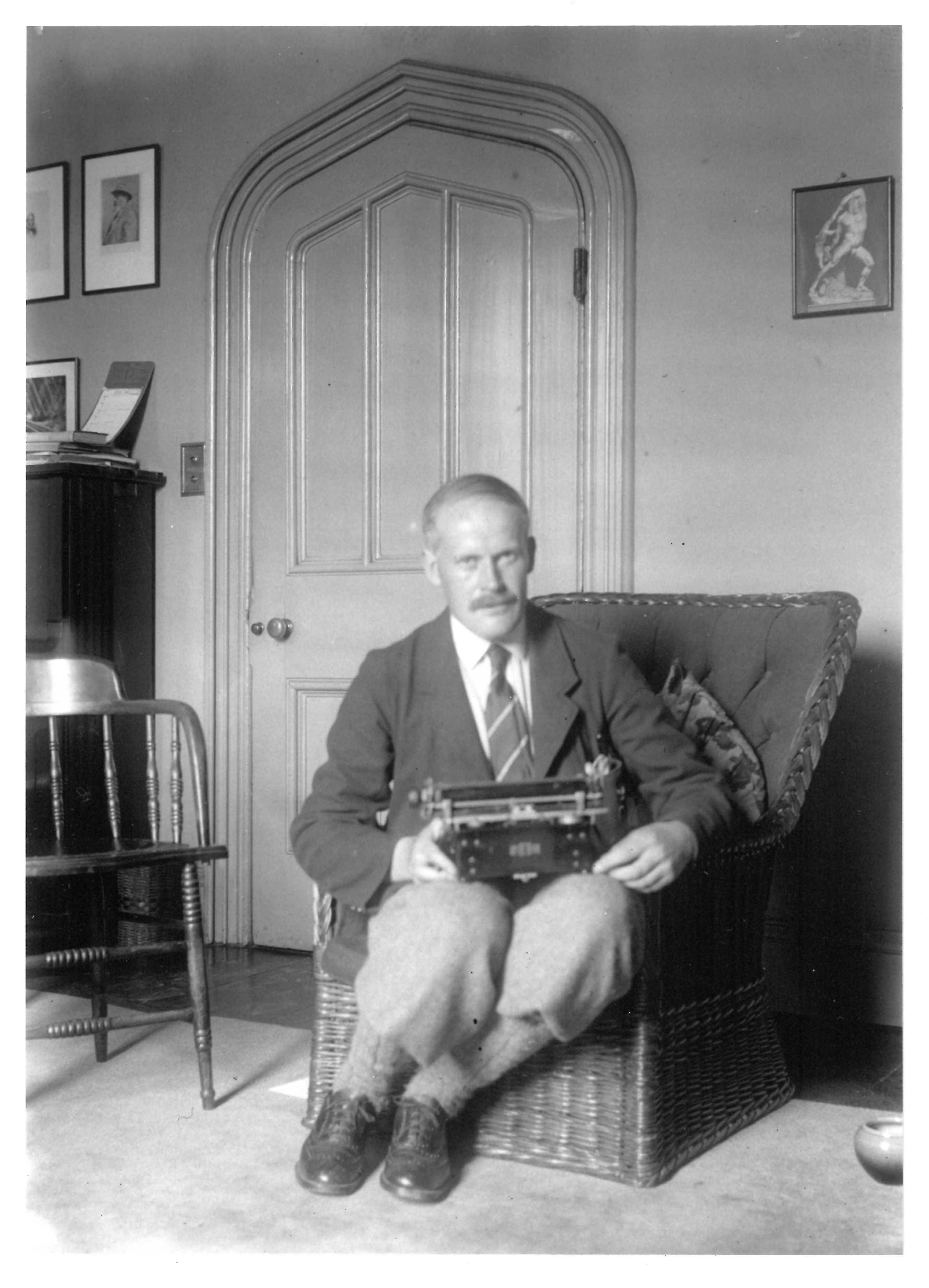} 
\caption{Sir Harold Jeffreys (1891--1989) in 1928. (Photographer unknown, included by permission of the Master and Fellows of St John's College, Cambridge)}
\end{figure}
Jeffreys calls his procedures ``significance tests" and surely means for this work to contrast with that of Fisher's. Even though Jeffreys does not mention Fisher directly, Jeffreys does allude to their earlier dispute by reaffirming that his probabilities express ``no opinion about the frequency of the truth of [the hypothesis] among any real or imaginary populations'' (presumably addressing one of \citeauthor{Fisher1934prob}'s \citeyear{Fisher1934prob} objections to Jeffreys's solution to the theory of errors) and that assigning equal probabilities to propositions ``is simply the formal way of saying that we do not know whether it is true or not of the actual populations under consideration at the moment'' \citep[p.\ 203]{Jeffreys1935}. 

Jeffreys strives to use the theory of probability to find a satisfactory rule to determine whether differences between observations should be considered ``significant,'' and he begins with a novel approach to test a difference of proportions as present in a contingency table. Jeffreys first posits the existence of two ``large, but not infinite, populations'' and that these ``have been sampled in respect to a certain property'' \citep[p.\ 203]{Jeffreys1935}. He continues,\footnote{To facilitate comparisons to Haldane's section we have translated Jeffreys's unconventional notation to a more modern form. See \citet[appendix D]{LyEtAlinpress} for details about translating Jeffreys's notation.}

\begin{quote}
One gives $[x_0]$ specimens with the property, $[y_0]$ without; the other gives $[x_1]$ and $[y_1]$ respectively. The question is, whether the difference between $[x_0/y_0]$ and $[x_1/y_1]$ gives any ground for inferring a difference between the corresponding ratios in the complete populations. Let us suppose that in the first population the fraction of the whole possessing the property is $[\theta_0]$, in the second $[\theta_1]$. Then we are really being asked whether $[\theta_0=\theta_1]$; and further, if $[\theta_0=\theta_1]$, what is the posterior probability distribution among values of $[\theta_0]$; but, if $[\theta_0\neq \theta_1]$, what is the distribution among values of $[\theta_0]$ and $[\theta_1]$. (p.\ 203)
\end{quote}
Take $\mathcal M_0$ to represent the proposition that $\theta_0=\theta_1$, so that $\mathcal M_1$ represents $\theta_0\neq \theta_1$. Jeffreys takes the prior probability of both $\mathcal M_0$ and $\mathcal M_1$ as $.5$. This problem can be restated so that $\mathcal M_0$ represents the difference between $\theta_0$ and $\theta_1$, namely, $\theta_0-\theta_1=0$. This is Jeffreys's point null hypothesis which is assigned a finite prior probability of $.5$. If the null is true, and $\theta_0=\theta_1$, then Jeffreys assigns $\theta_0$ a uniform prior distribution in the range $0-1$. 
The remaining half of the prior probability is assigned to $\mathcal M_1$, which specifies $\theta_0$ and $\theta_1$ have their probabilities ``uniformly and independently distributed'' in the range $0-1$ (p.\ 204). Thus,
$$\begin{array}{lcr}
\pi_0(\theta_0)&=&\mathcal U(0,1)\\
\pi_1(\theta_0)&=& \mathcal U(0,1)\\
\pi_1(\theta_1)&=&\mathcal U(0,1)
\end{array}$$ 
and $\pi_1(\theta_0,\theta_1)=\pi_1(\theta_0)\pi_1(\theta_1)$. 
Subsequently,
$$\begin{array}{lcl}
P(\mathcal M_0,\theta_0)&=&P(\mathcal M_0)\pi_0(\theta_0)\\ 
P(\mathcal M_1,\theta_0,\theta_1)&=&P(\mathcal M_1)\pi_1(\theta_0,\theta_1),
\end{array}$$
which follows from the product rule of probability, namely, 
$$P(p,q)=P(p)P(q\mid p).$$

Jeffreys then derives the likelihood functions for $D$ (i.e., the data, or compositions of the two samples) on the null and the alternative. Under the null hypothesis, $\mathcal M_0$, the probability of $D$ is
$$f(D \mid \theta_0,\mathcal M_0)=\frac{(x_0+y_0)!}{x_0!y_0!}
\frac{(x_1+y_1)!}{x_1!y_1!}
\theta_0^{x_0}(1-\theta_0)^{y_0}\,
\theta_0^{x_1}(1-\theta_0)^{y_1}.$$
Under the alternative hypothesis, $\mathcal M_1$, the probability of $D$ is
$$f(D \mid \theta_0,\theta_1,\mathcal M_1)=\frac{(x_0+y_0)!}{x_0!y_0!}
\frac{(x_1+y_1)!}{x_1!y_1!}
\theta_0^{x_0}(1-\theta_0)^{y_0}\,
\theta_1^{x_1}(1-\theta_1)^{y_1}.$$
In other words, the probability of both sets of data ($x_0+y_0$ and $x_1+y_1$), on $\mathcal M_0$ or $\mathcal M_1$, is equal to the product of their respective binomial likelihood functions and constants of proportionality. Since the prior distributions for $\theta_0$ and $\theta_1$ are uniform 
and the prior probabilities of $\mathcal M_0$ and $\mathcal M_1$ are equal, the posterior distributions of $\theta_0$ and $(\theta_0,\theta_1)$ are proportional to their respective likelihood functions above. Thus the posterior distributions for $\theta_0$ and $(\theta_0,\theta_1)$ under the null and alternative are, respectively,
$$\pi_0(\theta_0\mid D)\propto 
\theta_0^{x_0+x_1}(1-\theta_0)^{y_0+y_1},$$
and
$$\pi_1(\theta_0,\theta_1\mid D)\propto 
\theta_0^{x_0}(1-\theta_0)^{y_0}\,
\theta_1^{x_1}(1-\theta_1)^{y_1}.$$
Now Jeffreys has two contending posterior distributions, and to find the posterior probability of $\mathcal M_1$ he integrates $P(\mathcal M_0, \theta_0\mid D)$ with respect to $\theta_0$ and integrates $P(\mathcal M_1,\theta_0,\theta_1\mid D)$ with respect to both $\theta_0$ and $\theta_1$. Using the identity $\int_0^1\theta_0^{x_0}(1-\theta_0)^{y_0}\text{d}\theta_0=(x_0!y_0!)/(x_0+y_0+1)!$, he derives the relative posterior probabilities of $\mathcal M_0$ and $\mathcal M_1$. For $\mathcal M_0$ it is simply the previous identity with the additional terms of $x_1$ and $y_1$ added to the respective factorial terms,
$$P(\mathcal M_0 \mid D)\propto \frac{(x_0+x_1)!(y_0+y_1)!}{(x_0+x_1+y_0+y_1+1)!}. $$
To obtain the posterior probability of $\mathcal M_1$, the product of the integrals with respect to each of $\theta_0$ and $\theta_1$ is needed,
$$P(\mathcal M_1 \mid D)\propto \frac{x_0!y_0!}{(x_0+y_0+1)!} \frac{x_1!y_1!}{(x_1+y_1+1)!}.$$
Their ratio gives the posterior odds, $P(\mathcal M_0 \mid D)/P(\mathcal M_1 \mid D)$, and is the solution to Jeffreys's significance test. When this ratio is greater than $1$, the data support $\mathcal M_0$ over $\mathcal M_1$, and vice versa. 

Jeffreys ends his paper with a discussion of the implications of his approach.\footnote{In this paper Jeffreys gives many more derivations of new types of significance tests, but one example suffices to convey the general principle.} One crucial point is that if the prior odds are different from unity then the final calculations from Jeffreys's tests yield the marginal likelihoods of the two hypotheses, $P(D\mid \mathcal M_0)$ and $P(D\mid \mathcal M_1)$, whose ratio gives the Bayes factor. We quote at length,
\begin{quote}
We have in each case considered the existence and the non-existence of a real difference between the two quantities estimated as two equivalent alternatives, each with prior probability $1/2$. This is a common case, but not general. If however the prior probabilities are unequal the only difference is that the expression obtained for 
$[P(\mathcal M_0 \mid D)/P(\mathcal M_1 \mid D)]$ now represents $\left[\dfrac{P(\mathcal M_0 \mid D)}{P(\mathcal M_1 \mid D)} \bigg/
\dfrac{P(\mathcal M_0)}{P(\mathcal M_1)}\right]$[NB: the Bayes factor, which is the posterior odds divided by prior odds].
Thus if the estimated ratio exceeds 1, the proposition $[\mathcal M_0]$ is rendered more probable by the observations, and if it is less than 1, $[\mathcal M_0]$ is less probable than before. It still remains true that there is a critical value of the observed difference, such that smaller values reduce the probability of a real difference. The usual practice [NB: alluding to Fisher] is to say that a difference becomes significant at some rather arbitrary multiple of the standard error; the present method enables us to say what that value should be. If however the difference examined is one that previous considerations make unlikely to exist, then we are entitled to ask for a greater increase of the probability before we accept it, and therefore for a larger ratio of the difference to its standard error. (p.\ 221)
\end{quote}
Here Jeffreys has also laid out the benefits his tests possess over Fisher's tests: The critical standard error, where the ratio supports $\mathcal M_0$, is not fixed at an arbitrary value but is determined by the amount of data \citep[p.\ 205]{Jeffreys1935}; in Jeffreys's test, larger sample sizes increase the critical standard error and thus increase the barrier for `rejection' at a given threshold (note that this anticipates Lindley's paradox, \citealp{Lindley1957}). Furthermore, if the phenomenon has unfavorable prior odds against its existence we may reasonably require more evidence (i.e., a larger Bayes factor) before we are reasonably confident in its existence. That is to say, extraordinary claims require extraordinary evidence.

In a complementary paper, \citet{Jeffreys1936furthersig} expanded on the implications of this method: 

\begin{quote}
To put the matter in other words, if an observed difference is found to be on order [of one standard error from the null], then on the hypothesis that there is no real difference this is what would be expected; but if there was a real difference that might have been anywhere within a range $m$ it is a remarkable coincidence that it should have happened to be in just this particular stretch near zero. On the other hand if the observed difference is several times its standard error it is very unlikely to have occurred if there was no real difference, but it is as likely as ever to have occurred if there was a real difference. In this case beyond a certain value of $x$ [a distance from the null] the more remarkable coincidence is for the hypothesis of no real difference \dots The theory merely develops these elementary considerations quantitatively\dots (p.\ 417)
\end{quote}

In sum, the key developments in Jeffreys's significance tests are that a point null is assigned finite prior probability and that this point null is tested against a distributed (composite) alternative hypothesis. The posterior odds of the models are computed from the data, with the updating factor now known as the Bayes factor. 

\section{The Connection Between Jeffreys and Haldane}
We are now in a position to draw comparisons between the developments of Haldane and Jeffreys. 
The methods show a striking similarity in that they both set up competing models with orthogonal parameters, each with a finite prior model probability, and through the calculation of each model's marginal likelihood find posterior model probabilities and posterior distributions for the parameters. Where the methods differ is primarily in the focus of their inference. 
Haldane focuses on the overall mixture posterior distribution for the parameter, $\pi(\theta\mid D)$, marginalized across the competing models. In Haldane's example, this means to focus on estimating the cross-over rate parameter, using relevant real-world knowledge of the problem to construct a mixture probability distribution. It would have been but a short step for Haldane to find the ratio of the posterior and prior model odds as Jeffreys did, since the prior and posterior model probabilities are crucial in constructing the mixture distributions, but that aspect of the problem was not Haldane's primary interest. 

Jeffreys's focus is nearly the opposite of Haldane's. Jeffreys uses the models to instantiate competing scientific theories, and his focus is on making an inference about which theory is more probable.
In contrast to Haldane, Jeffreys isolates the value of the Bayes factor as the basis of scientific learning and statistical inference, and he takes the posterior distributions for the parameters within the models as a mere secondary interest. If one does happen to be interested in estimating a parameter in the presence of model uncertainty, Jeffreys recognizes that one should ideally form a mixture posterior distribution for the parameter (as Haldane does), but that ``nobody is likely to use it. In practice, for sheer convenience, we shall work with a single hypothesis and choose the most probable'' \citep[p.\ 222]{Jeffreys1935}.   

Clearly there was a great concurrence of thought between Haldane and Jeffreys during this time period. A natural question is whether the men knew of each others' work on the topic. In his paper, \citet{Haldane1932} makes no reference to any of Jeffreys's previous works (recall, Haldane's stated goal was to extend Fisher's method of maximum likelihood and discuss the use of non-uniform priors). Jeffreys's first book, \textit{Scientific Inference} \citep{Jeffreys1931}, came out a mere eight months before Haldane submitted his paper, and it was not widely read  by Jeffreys's contemporaries; around the time the revised first edition was released in 1937, Jeffreys remarked to Fisher in a letter (on June 5, 1937) that he ``should really have liked to scrap the whole lot and do it again, but at the present rate it looked as if the thing would take about 60 years to sell out'' \citep[p.\ 164]{bennett1990}.\footnote{Indeed, whereas 224 copies were sold in Great Britain and Ireland in 1931, only twenty copies in total were sold in 1932, sixty-nine in 1933, thirty-five in 1934, fourteen in 1935, and twenty-two in 1936 (Exhibits D399-400, St John's College Library, Papers of Sir Harold Jeffreys).} So it would be no surprise if Haldane had not come across Jeffreys's book. In fact, had Haldane known of \textit{Scientific Inference}, he would have surely recognized that Jeffreys had given the same topics a thorough conceptual treatment. For example, Haldane might have cited \citet{WrinchJeffreys1919}, who detached the theory of inverse probability from uniform distributions over a decade before; or Haldane might have recognized the complete form of the Haldane prior presented by \citet[p. 194]{Jeffreys1931}. Therefore, according to the evidence in the literature, one might reasonably conclude that by 1931/1932, Haldane did not know of Jeffreys's work. What about Jeffreys, is there evidence in the literature that he knew of Haldane's work while working on his significance tests?

According to \citet{Jeffreys1977}, in the footnote in a piece looking back on his career, he did recognize the similarity of his work to Haldane's:

\begin{quote}
The essential point [in solving Broad's quandary] is that when we consider a general law we are supposing that it may possibly be true, and we express this by concentrating a positive (non-zero) fraction of the initial probability in it. Before my work on significance tests the point had been made by J. B. S. Haldane (1932). (p.\ 95)
\end{quote}
Jeffreys also acknowledges Haldane in a footnote of his first edition of \textit{Theory of Probability} (\citeyear{Jeffreys1939}, retained in subsequent editions) when he remarks, ``This paper [\citep{Haldane1932}] contained the use of \dots the concentration of a finite fraction of the prior probability in a particular value, \textit{which later became the basis of my significance tests}''[emphasis added] (p.\ 114, footnote). So it is clear that Jeffreys, at least some time later, recognized the similarity of his and Haldane's thinking. 

Jeffreys's awareness of Haldane's work must go back even further. Jeffreys certainly knew of Haldane's work by 1932, because he wrote a critical commentary on Haldane's paper (which Jeffreys read at \textit{The Cambridge Philosophical Society} on November 7, 1932), pointing out, 
\textit{inter alia}, 
that Haldane's proposed prior distribution largely ``agrees with a previous guess'' of his, but only after a slight correction to its form \citep[p.\ 85]{Jeffreys1933}.\footnote{See \citet[p.\ 121--126]{howie2002} for more detail on the reactions to Haldane's paper by both Jeffreys and Fisher; \citet{Fisher1932} was particularly pointed in his commentary.} However, when publishing his seminal significance test papers in 1935 and 1936 \citep{Jeffreys1935,Jeffreys1936furthersig}, Jeffreys does not mention Haldane's work. It may appear to a reader of those papers as if Jeffreys's tests are entirely his own invention; indeed, they are the natural way to apply his early work with Wrinch on the philosophy of induction to quantitative tests of general laws. Perhaps Jeffreys simply forgot, or did not realize the significance of Haldane's paper on his own thinking at the time. But in another paper published at that time, \citet{Jeffreys1936criticisms}, in pointing out how he and Wrinch solved Broad's quandary, remarked,

\begin{quote}
but we [Wrinch and Jeffreys] pointed out that agreement with general belief is obtained if we take the prior probability of a simple law to be finite, whereas on the natural extension of the previous theory it is infinitesimal. Similarly for the case of sampling J.\ B.\ S.\ Haldane and I have pointed out that general laws can be established with reasonable probabilities if their prior probabilities are moderate and independent of the whole number of members of the class sampled. (p.\ 344)
\end{quote}
So it seems that Jeffreys did recognize the importance and similarity of Haldane's work in the very same years he was publishing his own work on significance tests (1935--1936). Why Jeffreys did not cite Haldane's work directly in connection with his own development of significance tests is not clear just from looking at the literature. There is, however, some potential clarity to be gained by examining the personal relationship between the two men.

Haldane and Jeffreys were both in Cambridge from 1922-1932 while working on these problems (Haldane at Trinity College and Jeffreys at St. John's), after which Haldane left for University College London. 
In an (unpublished) interview with George Barnard (recall we quoted Barnard above as being one of the few to recognize Haldane's innovative ``lump'' of prior probability), after Jeffreys (HJ) denies having known Fisher while they were both at Cambridge, Barnard (GB) asks,
\begin{quote}
GB: But Haldane was in Cambridge, was he?

HJ: Yes.

GB: Because he joined in a bit I think in some of the \dots [NB: interrupted]

HJ: Yes, well Haldane did anticipate some things of mine. I have a reference to him somewhere.

GB: But the contact was via the papers [NB: \citealp{Haldane1932, Jeffreys1933}] rather than direct personal contact. 

HJ: Well I knew Haldane, of course. Very well. 

GB: Oh, ah.

HJ: Can you imagine him being in India?

(Transcribed by AE from the audio-cassette in Exhibit H204, St John's College Library, Papers of Sir Harold Jeffreys)
\end{quote}
So it would seem that Jeffreys and Haldane knew each other personally very well in their time at Cambridge. We cannot be sure of the extent to which they knew of each others' professional work at that time, because we can find no surviving direct correspondence between Jeffreys and Haldane during the 1920s to early 1930s; we are unable to do more than speculate about what topics they might have discussed in their time together at Cambridge. However, there is some preserved correspondence from the 1940s where they discuss, among other topics, materialist versus realist philosophy, scientific inference, geophysics, and Marxist politics.\footnote{This correspondence is available online thanks to the UCL Wellcome Library: \url{http://goo.gl/9qHBF6}. Interestingly, near the end of this correspondence (undated, but from some time between 2 October 1942 and 29 March 1943) Haldane says to Jeffreys that he felt ``emboldened by your [Jeffreys's] kindness to my views on inverse probability.''} It stands to reason that Haldane and Jeffreys would have discussed similar topics in their time together at Cambridge.

This personal relationship opens up the possibility that Jeffreys and Haldane discussed their work in detail and consciously chose not to cite each other in their published works in the 1930s. Perhaps Haldane brought up the genetics problems he was working on and Jeffreys suggested the idea to Haldane to set up orthogonal models as he did, an application which Jeffreys %
thought of but %
had not yet published on. Haldane goes on to publish his paper, and both men, realizing the question of credit is somewhat murky, decide to ignore the matter. %
\ins{This would explain why Haldane never references these developments later in his career.} %
Admittedly, this account leaves a few points unresolved. %
\ins{It does not appear that Haldane used this method in any of his later empirical investigations; if Haldane and Jeffreys together developed this method in order to apply it to one of Haldane's genetics problems, why did Haldane never go on to actually apply it?}
And if Jeffreys had thought of how to apply this revolutionary method of inference, why was it not included in his book? Under this account, Jeffreys would have had to think of this development in the 
few months %
between when he wrote and published \textit{Scientific Inference} and when Haldane began to write his paper. Moreover, in his interview with Barnard and in his later published works, Jeffreys readily notes that Haldane anticipated some of his own ideas (although, it is not always entirely clear to which ideas Jeffreys refers). Did Jeffreys begin to feel guilty that he would be getting credit for ideas that Haldane helped develop?

Today one might be surprised to hear that two people could become good friends while potentially never discussing their work with each other; 
however, Jeffreys has a history of remaining unaware of his close probabilist contemporaries' work. It is well-known that  
Jeffreys and fellow Cambridge probabilist Frank Ramsey were good friends while at Cambridge and they never discussed their work with each other either. Ramsey was highly regarded in Cambridge at the time, and as Jeffreys recalls in his interview with Lindley, ``I knew Frank Ramsey well and visited him in his last illness but somehow or other neither of us knew that the other was working on probability theory'' (``Transcription of a Conversation between Sir Harold Jeffreys and Professor D.V. Lindley'', Exhibit A25, St John's College Library, Papers of Sir Harold Jeffreys).\footnote{Jeffreys was also unaware of the work of Italian probabilist Bruno de Finetti, another of his contemporary probabilists. In their interview, Lindley asks Jeffreys if he and de Finetti ever made contact. Jeffreys replies that, not only had he and de Finetti never been in contact, ``I've only just heard his name... I've never seen anything that he's done... I'm afraid I've just never come across him'' (``Transcription of a Conversation between Sir Harold Jeffreys and Professor D.V. Lindley'', Exhibit A25, St John's College Library, Papers of Sir Harold Jeffreys).}
When one realizes that the friendship between Jeffreys and Ramsey began with their shared interest in psychoanalysis \cite[p.\ 117]{howie2002}, perhaps it makes sense that they would not get around to discussing their work on such an obscure topic as probability theory. 

\section{Concluding Comments}
Around 1920, Dorothy Wrinch and Harold Jeffreys were the first to note that in order for induction to be possible it is essential that general laws must be assigned finite initial probability. This argument was revolutionary in that it went against their contemporaries' blind adherence to Laplace's principle of indifference. It is this brilliant insight of Wrinch and Jeffreys that forms the conceptual basis of the Bayes factor. Later Jeffreys would develop concrete Bayes factors in order to test a point null against a smoothly distributed alternative to evaluate whether the data justify changing the form of a general law. For these reasons, we believe that Wrinch and Jeffreys should together be credited as the progenitors of the concept of the Bayes factor, with Jeffreys the first to put the Bayes factor to use in real problems of inference.

However, our historical review suggests that in 1931 J.\ B.\ S.\ Haldane made an important intellectual advancement in the development of the Bayes factor. 
We speculate that it was the specific nature of the linkage problem in genetics that caused Haldane to serendipitously adopt a mixture prior comprising a point mass and smooth distribution; it does not appear as if Haldane derived his result from a principled philosophical stance on induction, in contrast to Jeffreys, but merely through a pragmatic attempt at utilizing non-standard (i.e., non-uniform) distributions with inverse probability. %
\ins{And yet, Haldane never went on to employ this method to any of his future applications, so we cannot discount the possibility that he was simply momentarily inspired to address the foundations of statistical inference---as a polymath is wont to do. Then, having lost interest, he never follows up on this work, thereby dooming his important development to obscurity. We may never know his true motivations.} %
Nevertheless, Haldane's work likely formed the impetus for Jeffreys to make his conceptual ideas concrete, leading to his thorough development of Bayes factors in the following years.

The personal relationship between Haldane and Jeffreys further complicates the story behind these developments. The two men were in close contact during the period when they developed their ideas, and the extent to which they knew of each other's work on essentially the same problem is unclear. Haldane and Jeffreys had closely converging ideas, as is seen by the similarity of their work in the 1930s, and both were statistical pioneers whose influence is still felt today. We hope this historical investigation will bring Haldane some well-deserved credit for his impact on the development of the Bayes factor.

\section*{Acknowledgements}
We are especially grateful to Kathryn McKee for helping us access the material in the Papers of Sir Harold Jeffreys special collection at the St John's College Library; the quoted material is included by permission of the Master and Fellows of St John's College, Cambridge. We would like to thank Anthony (A.\ W.\ F.) Edwards for keeping our Bernoullis straight, as well as looking up lecture lists from the 1930s to check on Turing and Jeffreys's potential interactions. We would like to thank Christian Robert and Stephen Stigler for helpful comments and suggested references. We also thank Alexander Ly for critical comments on an early draft and for fruitful discussions about the philosophy of Sir Harold Jeffreys. The first author (AE) is grateful to Rogier Kievit and Anne-Laura Van Harmelen for their hospitality during his visit to Cambridge. Finally, we thank the anonymous reviewers and editor for valuable comments and criticism.
This work was supported by the ERC grant ``Bayes or Bust'' and the National Science Foundation Graduate Research Fellowship Program \#DGE-1321846.

\bibliographystyle{ba}
\bibliography{Alexrefs.bib,addtlrefs.bib}

\end{document}